\documentclass{elsart}

\usepackage{graphicx}

\usepackage{amsmath}
\usepackage{amssymb}

\begin{document}

\begin{frontmatter}



\title{Memory effect in growing trees}


\author{Krzysztof Malarz\corauthref{km}}
\ead{malarz@agh.edu.pl}
\corauth[km]{Corresponding author. Fax: +48 12 6340010.}
\and
\author{Krzysztof Ku{\l}akowski}

\address{Department of Applied Computer Science,
Faculty of Physics and Nuclear Techniques,
AGH University of Science and Technology\\
al. Mickiewicza 30, PL-30059 Krak\'ow, Poland}

\begin{abstract}
We show that the structure of a growing tree preserves an information on the
shape of an initial graph.
For the exponential trees, evidence of this kind of memory is provided by means of the iterative equations, derived for the moments of the node-node distance
distribution.
Numerical calculations confirm the result and allow to extend the conclusion to the Barab\'asi--Albert scale-free trees.
The memory effect almost disappears, if subsequent nodes are connected to the network with more than one link.
\end{abstract}

\end{frontmatter}


\section{Introduction}
The problem of growing trees belongs to larger class of problems of
evolving networks --- a new area with many interdisciplinary applications, from
biology and computational science to linguistics \cite{ab1,drm,nwm}. In
statistical mechanics, we often investigate the state of thermodynamic
equilibrium, which is unique and therefore it cannot preserve any information. 
However, in other sciences memory on past states is an essential ingredient
of the system. Here we are interested in search how the structure
of the origin of a tree, i.e. of a graph from which the tree is constructed,
influences the overall characteristics of the growing system.

A network containing $N$ nodes is fully characterized by its connectivity matrix $\mathbf{C}$: $c_N(i,j)=1$ if the nodes $i,j$ are linked together, and $c_N(i,j)=0$ elsewhere.
More convenient but somewhat redundant is the distance matrix $\mathbf{S}$, where the matrix element $s_N(i,j)$ is the number of links along the shortest path from $i$ to $j$.
It is often simpler to describe a network statistically.
A local characteristics of a network includes the degree distribution, i.e. the probability that a node is linked to a given number $k$ of neighbors.
A global characteristics includes the node-node distance distribution.
Whereas the former can be treated as complete only conditionally \cite{drm2}, a few is known on the latter.
Recent progress of knowledge on the mean node-node distance $d\equiv [\langle s_N(i,j)\rangle]$ is due to applications of equilibrium statistical mechanics, scaling hypotheses and/or assumptions of lack of correlations between nodes
\cite{burda,dms,hol2,bia}.
Here, $\langle \cdots \rangle$ denotes an average over $N(N-1)$ non-diagonal matrix elements and $[ \cdots ]$ is an average over different matrices, i.e. different graphs.

By growing we mean adding subsequent nodes to an already existing graph.
When each node is added with one link only $(m=1)$, a tree --- a compact graph without loops and without multiple edges --- is formed.
In trees, a path between each two nodes is unique, and it cannot be changed during the growth process.
When a node is added, the node-node distance matrix $\mathbf{S}$ is increased by one column and one row.
Once the matrix elements are formed, they do not change their values.
However, if nodes are added with two or more links $(m>1)$, a kind of shortcuts are formed and some node-node distances may be shortened.

The main goal of this work is to demonstrate, that the node-node
distance distribution of a growing tree preserves an information on the
structure of the initial tree, from which it is formed.

Below we deal with two kinds of growing trees, which differ in the degree distribution.
Let us consider the linking of new nodes to randomly selected nodes.
When the selection is made without any preference, we obtain a so-called exponential tree.
In this case, the degree distribution $P(k)=2^{-k}$, where $k$ is the number of links of a node.
Nodes can be selected also with some preference with respect to their degree.
If the linking probability is proportional to the degree $k$, we obtain the scale-free or Barab\'asi--Albert networks \cite{r1}.
In this case, $P(k)\propto k^{-\gamma}$, with $\gamma=3$ \cite{ab1,drm,nwm}.

To achieve our goal, the simplest method is to calculate the mean node-node distance $d(N)$ for trees of $N$ nodes, the formation of which has started from two different trees with four nodes.
This is done in the next section with iterative equations, which has been derived recently for the exponential trees \cite{mckk}.
In Section \ref{sec_algorithm}, the growth algorithms are introduced, basing on an evolution of the distance matrix.
In Section \ref{sec_results}, numerical results are presented for the exponential trees and the Barab\'asi--Albert scale-free trees.
We show also that the memory on the ancestral network is much reduced, if the trees are substituted by graphs with cyclic paths, i.e. with $m>1$.
The last section is devoted to discussion.

\section{Weights of exponential trees}
\begin{figure}
\begin{center}
\includegraphics[width=.7\textwidth]{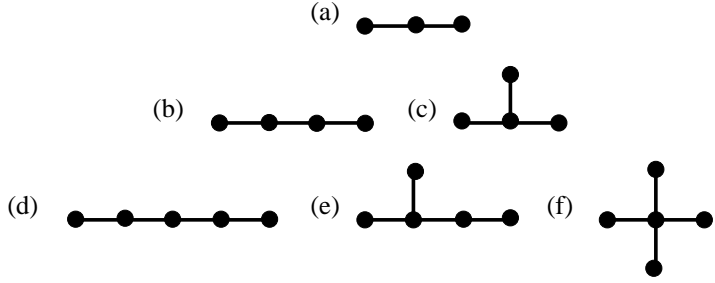}
\caption{Examples of trees. The Z-like chain (b) and the Y-like star (c) are the ancestors of the two ``families'' of growing networks described in the text.}
\label{fig_trees}
\end{center}
\end{figure}

Consider the probability that a tree of a given structure is grown. Trees are
different if there is no one-to-one correspondence between their pairs of
linked nodes \cite{wilsn}. Let us denote the number of different trees with
$N$ nodes by $K(N)$. It is easy to check by inspection, that $K(2)=K(3)=1$ and
$K(4)=2$. As $K(3)=1$, the probability --- or weight --- of the tree of three nodes
(Fig. \ref{fig_trees}(a)) must be one.
An exponential tree of four nodes can be formed by linking a new (fourth) node either to one of two end nodes, or to the central one.
Then, the probability of a chain of nodes (Fig. \ref{fig_trees}(b)) is $2/3$, and the probability of a star-like-tree (Fig. \ref{fig_trees}(c)) is $1/3$.
From the chain, a longer chain (Fig. \ref{fig_trees}(d)) can be produced in two ways, then its weight is $2/3 \cdot 2/4=1/3$.
From the star, another star (Fig. \ref{fig_trees}(f)) can appear with the probability $1/3\cdot 1/4=1/12$.
The remaining tree (Fig. \ref{fig_trees}(e)) can be formed from either the chain or the star, then its weight is $2/3\cdot 2/4+1/3\cdot 3/4=7/12$.
We note that in the case of the scale-free trees, the weights of the trees presented in Fig. \ref{fig_trees} are: 1, 1/2, 1/2, 1/6, 7/12 and 1/4, respectively.
This is a simple demonstration, that the weights of trees in two different classes are different.

Any possible tree can be formed from a tree of three nodes (Fig. \ref{fig_trees}(a)).
The way to form chains and stars is unique and then, their weights are relatively small.
Example giving, the weight of an exponential star of $N$ nodes is \mbox{$2/(N-1)!$}.
We could eliminate stars, if we develop trees from the chain shown in Fig. \ref{fig_trees}(b).
Seemingly, the weights of other trees should not be changed much, but all of them are influenced by the lack of the stars.
Example giving, in this case the tree shown in Fig. \ref{fig_trees}(e) can be formed in one unique way.
As a consequence, the whole distribution of weights is rebuilt.  
With the iterative equations derived recently \cite{mckk}, we can calculate the mean distance $d$ and the mean square of distances $e \equiv [\langle s_N^2(i,j)\rangle ]$ for two ``families'' of trees.
One is formed from the chain-like tree shown in Fig. \ref{fig_trees}(b) and labeled as ``Z'', and another --- from the star-like tree presented in Fig. \ref{fig_trees}(c) and marked as ``Y''.
Then, the first ``family'' does not contain stars, and the second one does not contain chains.
The equations are:   
\begin{subequations}
\label{eq_iter} 
\begin{equation} 
\label{eq_iter_d} 
d(N+1)=\frac{(N+2)(N-1)}{N(N+1)}d(N)+\frac{2}{N+1},
\end{equation} 
and
\begin{equation} 
\label{eq_iter_e} 
e(N+1)=\frac{(N+2)(N-1)}{N(N+1)}e(N)+\frac{4(N-1)}{N(N+1)}d(N)+\frac{2}{N+1}.
\end{equation} 
\end{subequations}
The information on the initial trees is encoded in the initial values of $d(4)$ and $e(4)$.
It is easy to check, that for the chain $d^Z(4)=5/3$, $e^Z(4)=10/3$ and for the star $d^Y(4)=3/2$, $e^Y(4)=5/2$.

Similar method has been used in \cite{kulkarni00,szabo02}.
The difference is that here, the Eqs. \eqref{eq_iter} are exact, but they apply only to the exponential trees.

\section{Numerical algorithm}
\label{sec_algorithm}
Two initial trees with four nodes (the chain and the star) are represented
in the computer memory as two distance matrices $\mathbf{S}(Z)$ and $\mathbf{S}(Y)$.
The starting point are two matrices for two trees of four nodes:
\[
\mathbf{S}_4(Z)=
\begin{pmatrix}
0&1&2&3\\
1&0&1&2\\
2&1&0&1\\
3&2&1&0\\
\end{pmatrix}
\text{ and }
\mathbf{S}_4(Y)=
\begin{pmatrix}
0&1&2&2\\
1&0&1&1\\
2&1&0&2\\
2&1&2&0\\
\end{pmatrix}
\]
for the chain and the star, respectively.

Selecting a node to link a new node is equivalent to select a number $q$ of column/row of the matrix.
Then the matrix is supplemented by new column and row, which are copies of the $q$-th column/row but with all elements incremented by one
\begin{subequations}
\begin{equation}
\label{eq_2a}
\forall ~ 1\le i \le N: s_{N+1}(N+1,i)=s_{N+1}(i,N+1)=s_N(q,i)+1,
\end{equation}
and obviously
\begin{equation}
s_{N+1}(N+1,N+1)=0.
\end{equation}
\end{subequations}

The Eq. \eqref{eq_2a} served in the derivation of the iterative formulas \eqref{eq_iter} \cite{mckk}.

The same numerical technique is applied also to the case of the Barab\'asi--Albert scale-free trees.
The only difference is that in this case, the node $q$ is selected with preference of the number of its pre-existing links.
Namely, 
\[
P(q)=k(q)/\sum_{i=1}^N k(i),
\]
where $k(i)$ is the number $k$ of links of $i$-th node.
Additional matrix $\mathbf{r}(i)$ contains the indices of row of the distance matrix $\mathbf{S}$ where ``1'' is encountered.
Each case $s_N(i,j)=1$ indicates a link between nodes $i$ and $j$.
The matrix $\mathbf{r}(i)$ is useful to select nodes of given degree for the scale-free trees and graphs, according to the so-called Kert\'esz algorithm \cite{stauffer-pc}.

Further, the same technique is applied to simple graphs, where new nodes are attached to previously existing ones by $m=2$ links.
Then, cyclic paths  are possible and the distance matrix $\mathbf{S}$ is to be rebuilt when adding each node.
The algorithm is as follows: Let us suppose that $(N+1)$-th node is added to existing nodes $p$ and $q\ne p$.
Then
\begin{subequations}
\label{eq_graphs}
\begin{equation}
\label{eq_graphs_a}
\begin{split}
\forall ~ 1\le i,j\le N: s_{N+1}(i,j)=\\
=\min\big(s_N(i,j),s_N(i,p)+2+s_N(q,j),s_N(i,q)+2+s_N(p,j)\big).
\end{split}
\end{equation}
For new, $(N+1)$-th, column/row
\begin{equation}
\label{eq_graphs_b}
\forall ~ 1\le i\le N: s_{N+1}(N+1,i)=s_{N+1}(i,N+1)=\min\big(s_N(p,i),s_N(q,i)\big)+1,
\end{equation}
and again for the diagonal element
\begin{equation}
\label{eq_graphs_c}
s_{N+1}(N+1,N+1)=0.
\end{equation}
\end{subequations}

\begin{figure}
\begin{center}
\includegraphics[width=.8\textwidth]{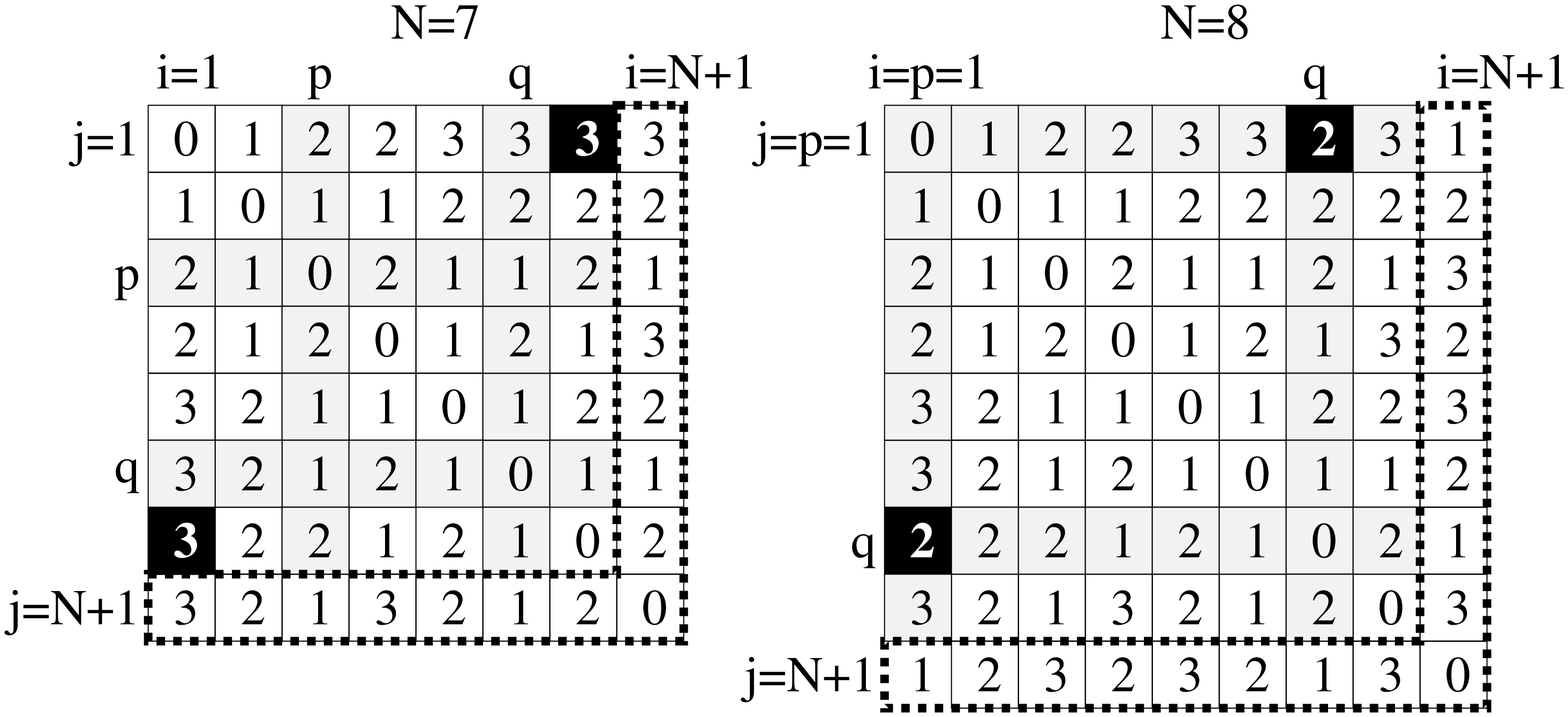}
\caption{Construction of the distance matrix $\mathbf{S}$ in the case of growing graphs $(m=2)$.
The gray sites show randomly chosen columns/rows (nodes to which new node will be attached).
The black sites show matrix elements which are reevaluated from Eq. \eqref{eq_graphs_a} due to newly created shortcuts.
The last columns/rows are constructed according Eqs. \eqref{eq_graphs_b} and \eqref{eq_graphs_c}.
Starting with the Y-like star new nodes were subsequently added to nodes $(p,q)=\big((3,4), (3,5), (4,6), (3,6), (1,7)\big)$.
}
\label{fig_matrix}
\end{center}
\end{figure}

One step of construction of the matrix $\mathbf{S}$ for simple graphs $(m=2)$ is presented in Fig. \ref{fig_matrix}.
An example of the construction $\mathbf{S}$ for trees $(m=1)$ is given in \cite{mckk}.

\section{Results of calculations}
\label{sec_results}

\begin{figure}
\begin{center}
\includegraphics[width=0.45\textwidth]{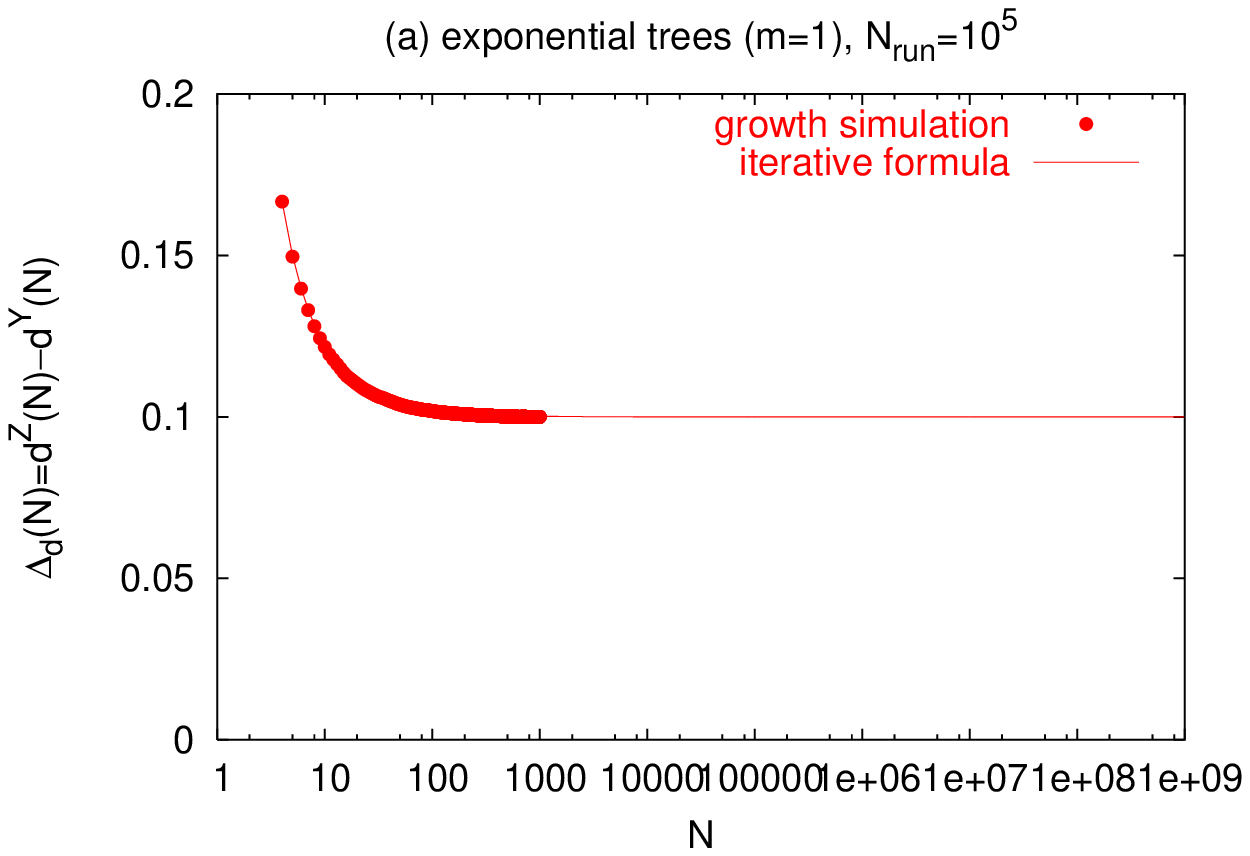}
\includegraphics[width=0.45\textwidth]{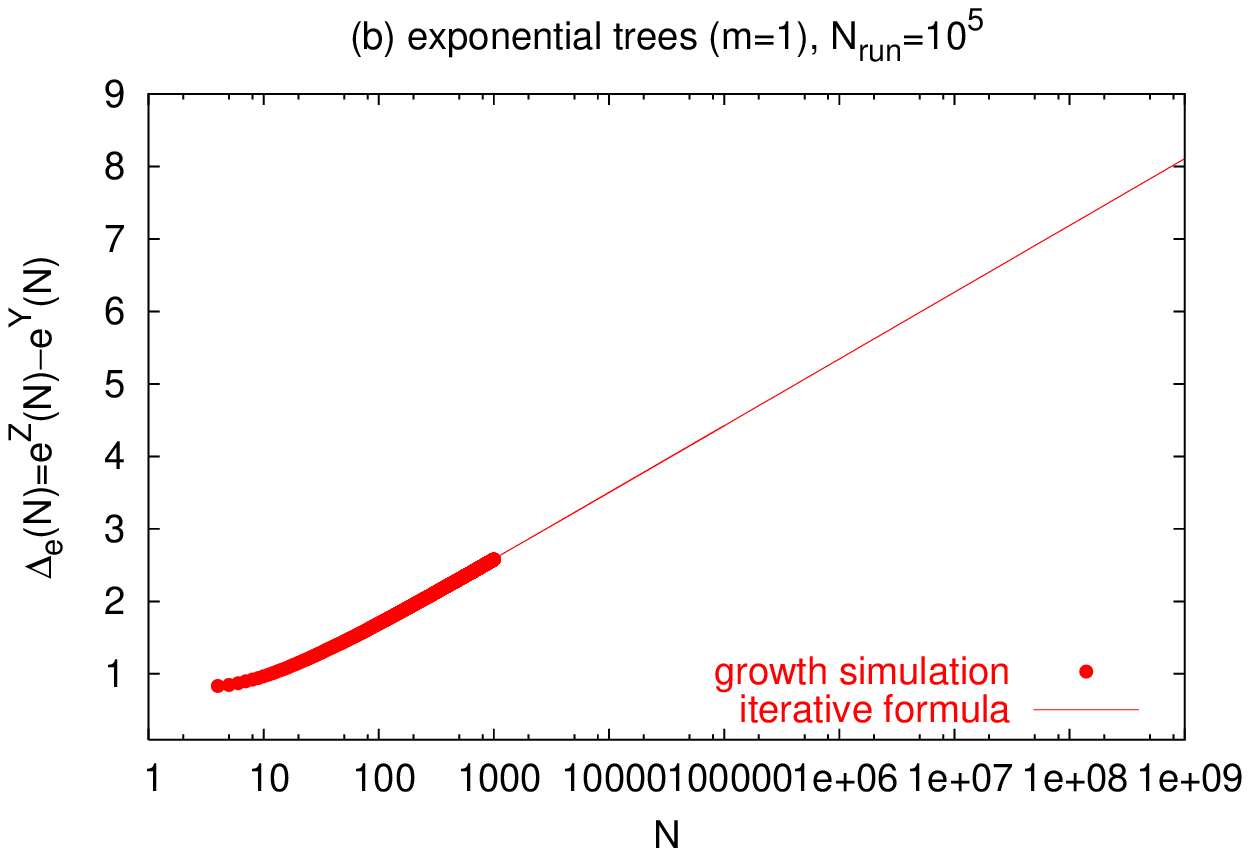}
\caption{The function (a) $\Delta_d(N)$ and (b) $\Delta_e(N)$ for {\em exponential trees} obtained with iterative formula \eqref{eq_iter} as well as from the direct growth simulations. The results of simulations are averaged over $N_{\text{run}}=10^5$ independent growths.}
\label{fig_met_exp}
\end{center}
\end{figure}

\begin{figure}
\begin{center}
\includegraphics[width=0.45\textwidth]{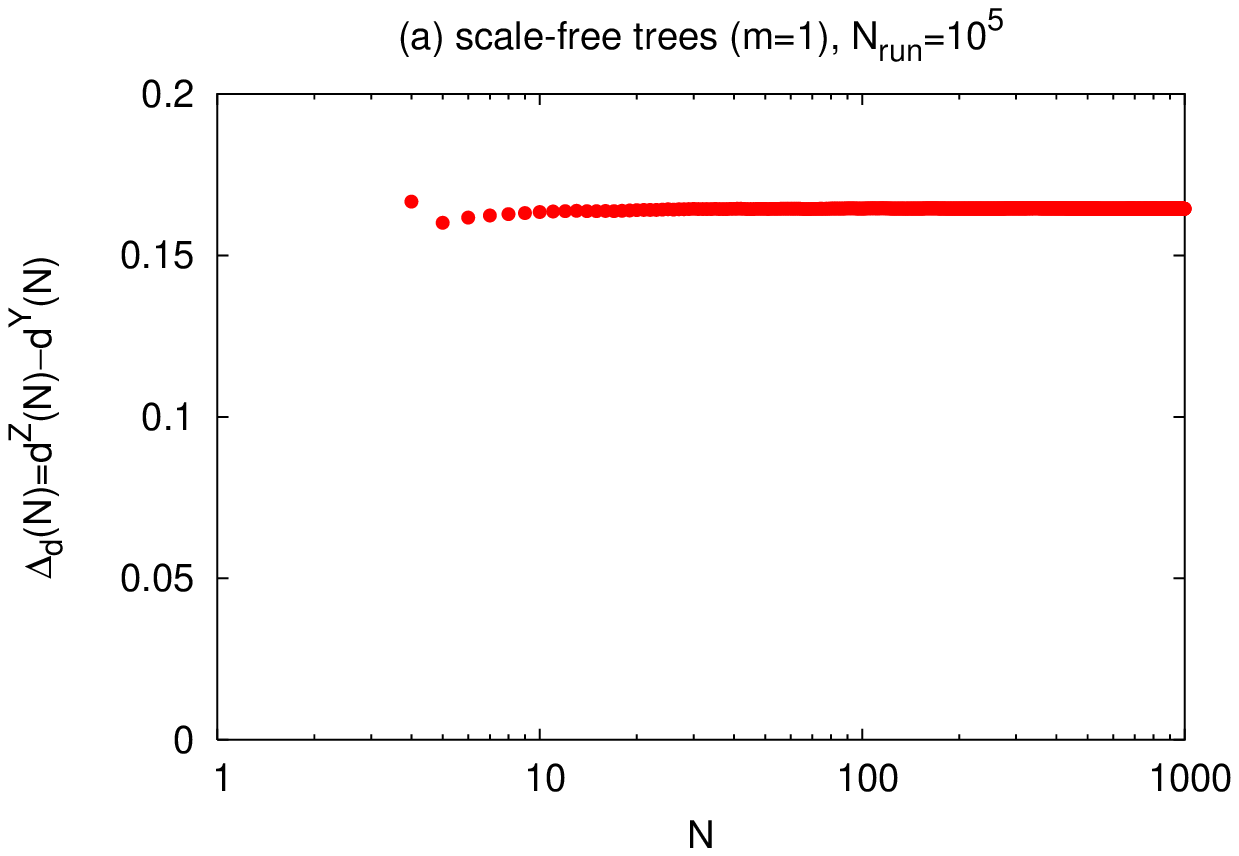}
\includegraphics[width=0.45\textwidth]{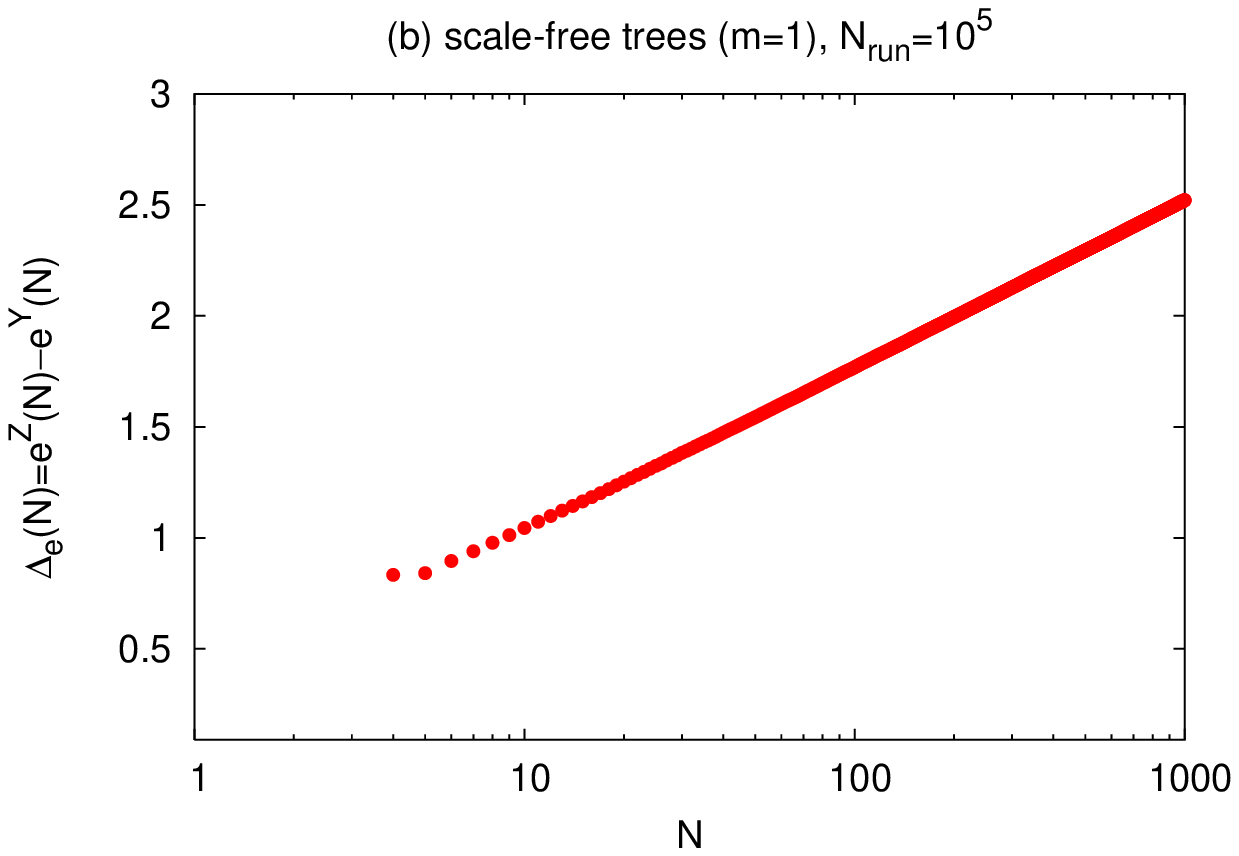}
\caption{The function (a) $\Delta_d(N)$ and (b) $\Delta_e(N)$ for {\em scale-free trees} obtained from the growth simulations. The results are averaged over $N_{\text{run}}=10^5$ independent growths.}
\label{fig_met_ab}
\end{center}
\end{figure}

In Figs. \ref{fig_met_exp} and \ref{fig_met_ab} the dependences (a) $\Delta_d(N)\equiv d^Z(N)-d^Y(N)$ and (b) $\Delta_e(N)\equiv e^Z(N)-e^Y(N)$ obtained from growth simulations are presented, for exponential trees and for scale-free trees, respectively.
The results of simulations are averaged over $N_{\text{run}}=10^5$ independent growths.
For both kinds of trees the difference in average node-node distance $\Delta_d$ tends to the constant value during the growth process.
For higher moments the effect is even stronger and $\Delta_e$ increases with the tree size $N$.
In Fig. \ref{fig_met_exp} we give also the results for $\Delta_d(N)$ and $\Delta_e(N)$ calculated with Eq. \eqref{eq_iter} for exponential trees of $N=10^9$ nodes.
The fact that $\Delta_d$ and $\Delta_e$ do not decrease with $N$ means that growing structures preserve the memory on their initial shapes.

\begin{figure}
\begin{center}
\includegraphics[width=0.45\textwidth]{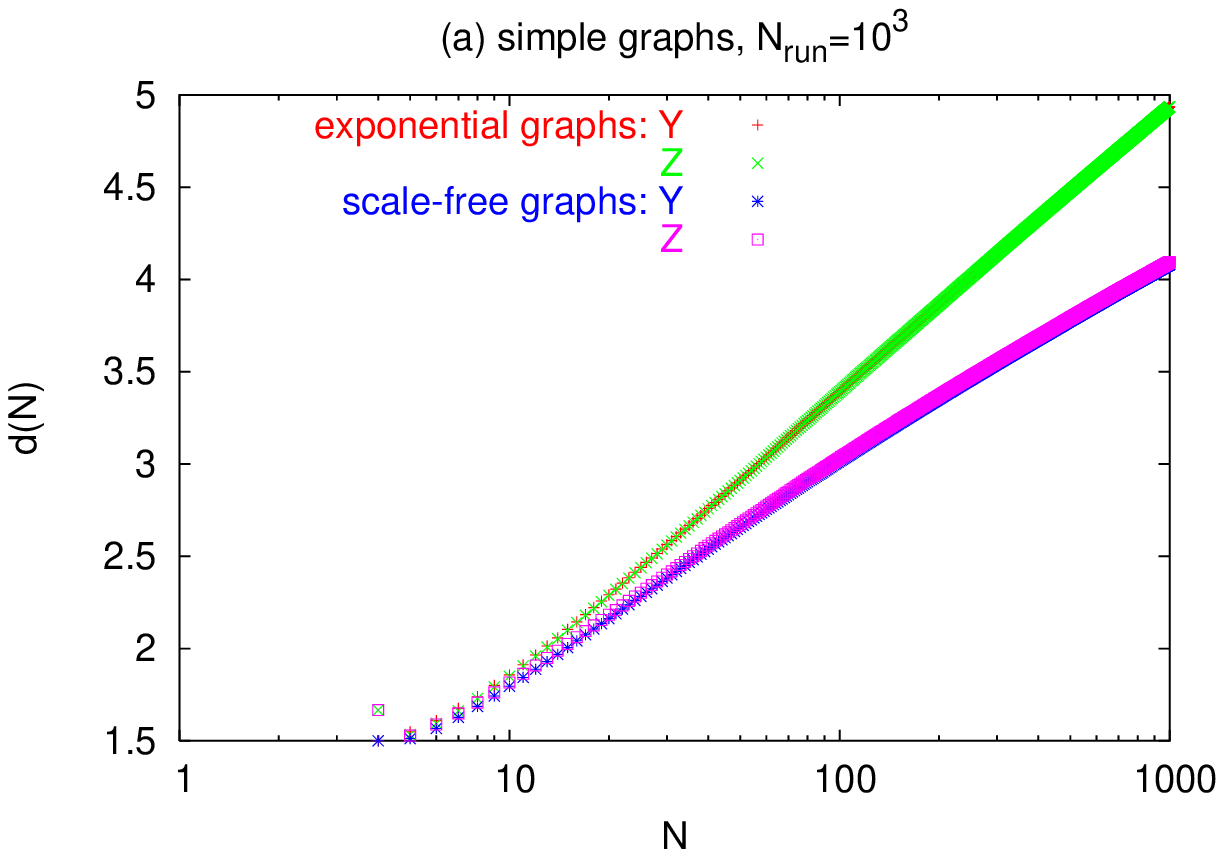}
\includegraphics[width=0.45\textwidth]{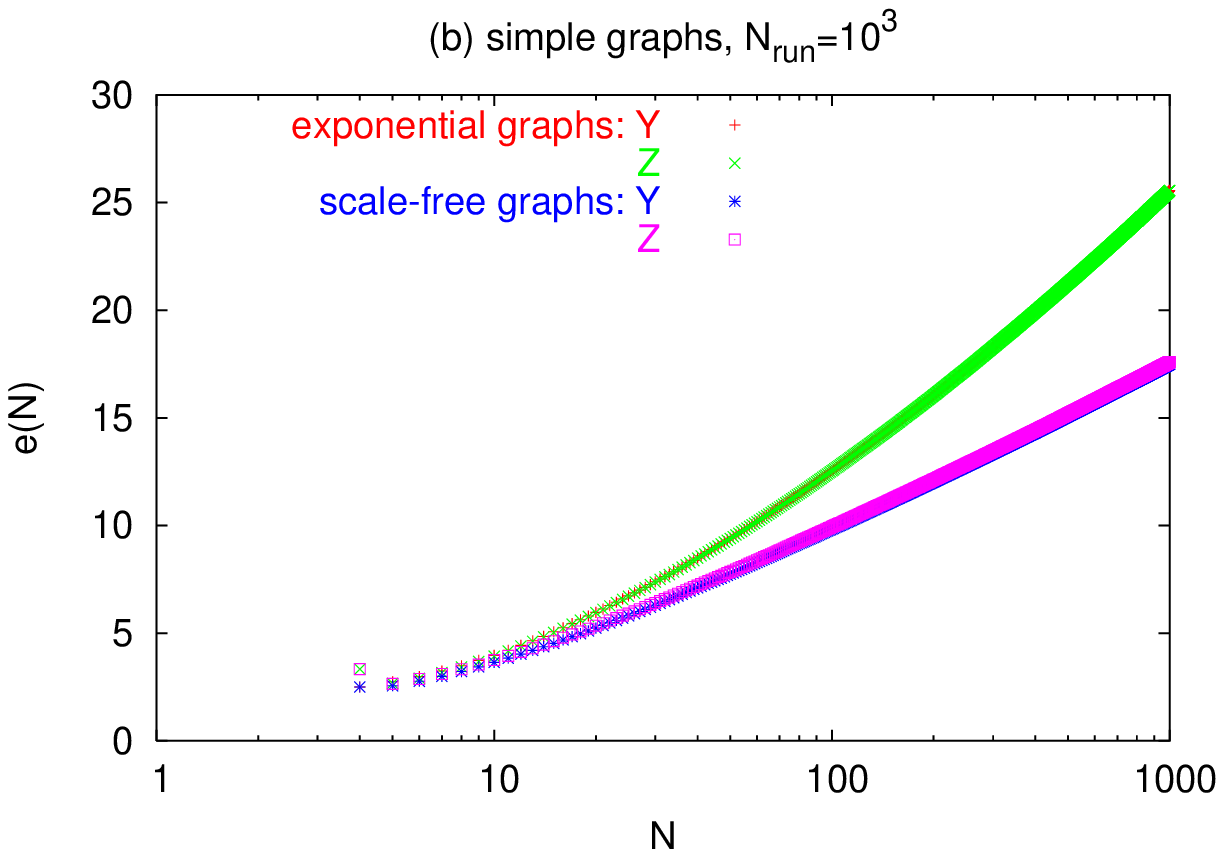}
\caption{The function (a) $d(N)$ and (b) $e(N)$ for the exponential and scale-free 
{\em graphs} and different initial configurations obtained from the growth simulations.
The results are averaged over $N_{\text{run}}=10^3$ independent growths.
The dependence on the initial configuration is not visible in the scale of the plot.}
\label{fig_dn}
\end{center}
\end{figure}

\begin{figure}
\begin{center}
\includegraphics[width=0.45\textwidth]{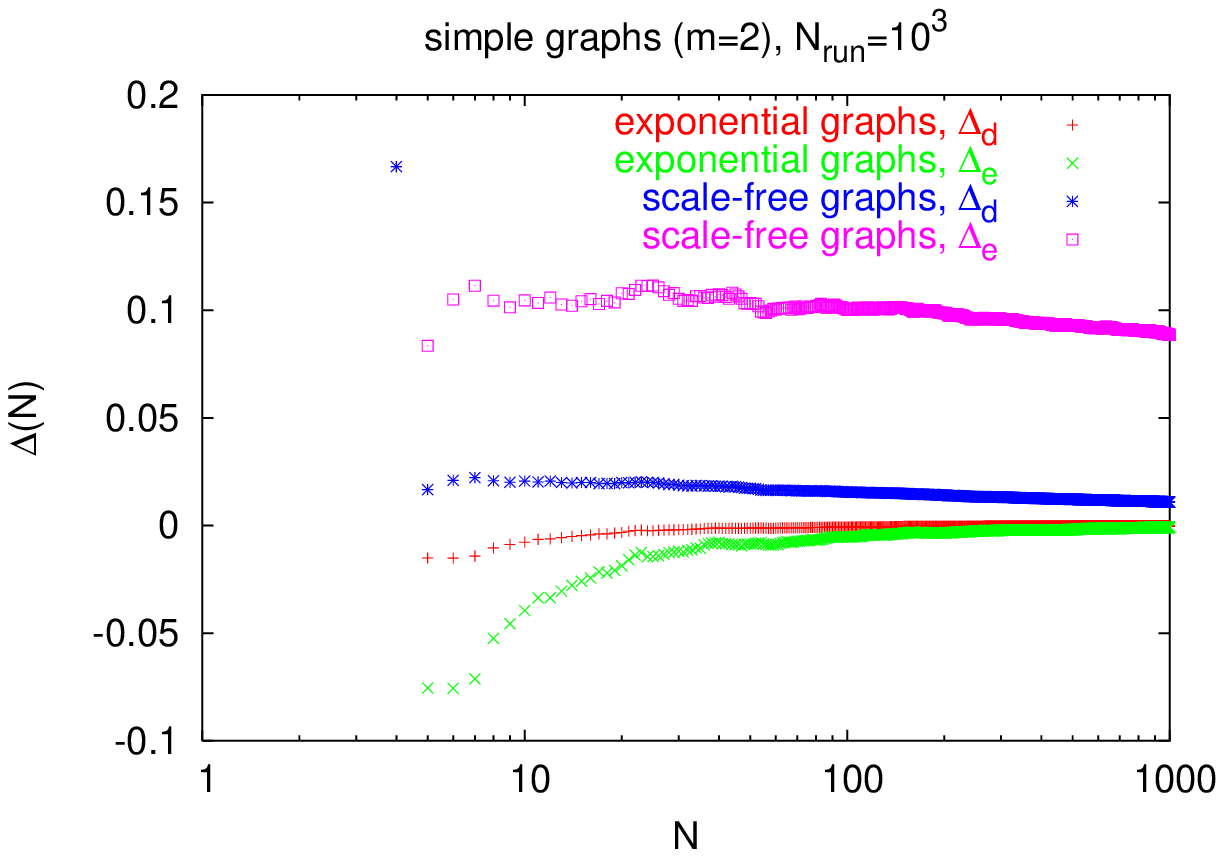}
\caption{The function $\Delta_d(N)$ and $\Delta_e(N)$ for the exponential and the scale-free 
{\em graphs} obtained from the growth simulations.
The results are averaged over $N_{\text{run}}=10^3$ independent growths.}
\label{fig_meg}
\end{center}
\end{figure}

In the case of simple graphs ($m=2$), the distance matrix $\mathbf{S}$ must be reevaluated, what makes the time of the calculation substantially larger.
The results for graphs are averaged only over one thousand of independent growths.
The curves $d(N)$ and $e(N)$ for both kind of simple graphs are shown in Fig. \ref{fig_dn}.
The linear fits for $100\le N\le 1000$ are $d(N)=0.672\ln(N)+0.299$ and $d(N)=0.462\ln(N)+0.888$ for the exponential graphs and the scale-free graphs, respectively.
The functions $\Delta_d(N)$ and $\Delta_e(N)$ for both kind of evolving graphs are shown in Fig. \ref{fig_meg}.

For the scale-free graphs, we observe some small memory effect, which manifests as a constant mutual shift of the plots $e(N)$ vs. $\ln(N)$.
In this case it is not clear if the effect vanishes or not, when $N$ tends to infinity.


\section{Discussion}
In the case of the exponential trees, the results of the simulations agree well 
with the curves obtained from the iteration equations.
This fact supports the reliability of the numerical equation for the scale-free trees 
and the graphs with $m=2$, where we have no analytical calculations.

Main result of this work is, that the node-node distance distribution in a growing 
tree depends on its initial structure.
Our calculations indicate, that both the average distance $d$ and its second moment 
$e$ in trees display this kind of memory.
The information is encoded in the constant $c_1$ in the expression $d=2\ln(N)+c_1$.
The constant $c_1$ varies by about 0.109 and 0.164, when we
change the shape of the initial tree of four nodes from the Y-like star to the Z-like 
chain for the exponential and scale-free trees, respectively.
In the second moment $e=4\ln^2(N)+c_2\ln(N)+c_3$,
it is the constant $c_2$ which depends on the initial shape. This is true both
for the exponential and the scale-free trees.

The memory effect is much reduced or even disappears in the case when new nodes are linked to
the network by at least two edges. In this case, the distance matrix $\mathbf{S}$ is
rebuilt by new edges which can shorten distances between initially far nodes
by providing new paths between them.

Concluding, we have demonstrated that the growing trees carry an information on
their initial geometrical structure.
The validity of this result relies not only on pure geometry, but also 
on a particular application of the graph theory. As remarked in the Introduction,
the list of examples of such applications is quite rich \cite{nwm}.  If 
the considered network is due to the citation index, we trace the flow of a new idea
from one paper to another. We see that around some
seminal papers, networks of citations are formed, as it happens in the case of
Ref. \cite{r1}. Sometimes there are two or more seminal papers, and then the
shape of the network depends on their clarity, ease of mathematical
formulation and individual preferences of the readership, formed in personal
contacts. If new results spread just by reading papers, it spreads slowly:
somebody reads it, tells to a friend, the friend's student is asked to calculate
a similar problem. The tree is `chain-like'. On the contrary, each conference makes 
the tree of spreading of ideas to be more `star-like', where the possible sources 
of getting new information are multiple and efficient.

\bigskip 

\ack
The authors are grateful to Prof. Kazimierz R\'o\.za\'nski for valuable help and to 
Mr. Pawe{\l} Ku{\l}akowski for comments on manuscript.
The numerical calculations were carried out in ACK--CYFRONET--AGH.
The machine time on SGI~2800 is financed by the Polish State Committee for Scientific 
Research (KBN) under grant No. KBN/\-SGI2800/\-AGH/\-018/\-2003.




\end{document}